\documentclass[12pt,preprint]{aastex}
\usepackage{natbib}
\bibpunct{(}{)}{;}{a}{}{,}

\usepackage{graphicx}

\usepackage{color,ulem}
\newcommand{\SUZAKU}{{\it Suzaku}}
\newcommand{\XMM}{{\it XMM-Newton}}
\newcommand{\INTEGRAL}{{\it INTEGRAL}}

\newcommand{\CHANDRA}{{\it Chandra}}
\newcommand{\Suzaku}{{\it Suzaku}}

\newcommand{\UNITCPS}{{\rm counts~s$^{-1}$}}
\newcommand{\UNITNH}{{\rm cm$^{-2}$}}
\newcommand{\ARCMIN}{{$'$}}
\newcommand{\DEGREE}{{$^{\circ}$}}
\newcommand{\NH}{{\it N$_{\rm H}$}}
\newcommand{\KT}{{\it kT}}
\newcommand{\EM}{{\it E.M.}}
\newcommand{\FOV}{{\it fov}}

\begin{document}

\title{High-Energy Properties of the Enigmatic Be Star $\gamma$
  Cassiopeiae}

\author{C. R. Shrader,\altaffilmark{1, 2}
  K. Hamaguchi,\altaffilmark{1, 3}, \& S. J. Sturner \altaffilmark{1,
    3}, L.M. Oskinova,\altaffilmark{4}, T. Almeyda,\altaffilmark{5},
  R. Petre\altaffilmark{1}}

\altaffiltext{1}{Astrophysics Science Division, NASA/GSFC, Greenbelt,
  MD 20771, Chris.R.Shrader@nasa.gov}

\altaffiltext{2}{Universities Space Research Association, 7178
  Columbia Gateway Drive Columbia, MD 21046 }

\altaffiltext{3}{Department of Physics, University of Maryland,
  Baltimore County, 1000 Hilltop Circle, Baltimore, MD 21250}

\altaffiltext{4}{Institute for Physics and Astronomy, University 
 Potsdam, 14476 Potsdam, Germany}

\altaffiltext{5}{Physics Department, Rochester Institute of 
 Technology, 54 Lomb Memorial Drive, Rochester, NY, 14623 }




\begin{abstract}

We present the results of a broad-band X-ray study of the enigmatic Be
star Gamma Cassiopeiae (herein $\gamma$ Cas) based on observations
made with both the {\it Suzaku} and {\it INTEGRAL} observatories.
$\gamma$~Cas has long been recognized as the prototypical example of a
small subclass of Be stars with moderately strong X-ray emission
dominated by a hot thermal component in the 0.5-12 keV energy range
($L_{\rm x} \approx 10^{32} - 10^{33}\ {\rm erg\ s^{-1}}$). This
places them at the high end of the known luminosity distribution for
stellar emission, but several orders of magnitude below typical
accretion powered {\bf Be} X-ray binaries.  The {\it INTEGRAL}
observations spanned an 8 year baseline and represent the deepest
measurement to date at energies above $\sim 50$ keV.  We find that the
{\it INTEGRAL} data are consistent within statistics to a constant
intensity source above 20 keV, with emission extending up to
$\sim$100~keV and that searches for all of the previously reported
periodicities of the system at lower energies led to null results. We
further find that our combined {\it Suzaku} and {\it INTEGRAL}
spectrum, which we suggest is the most accurate broad-band X-ray
measurement of $\gamma$ Cas to date, is fitted extremely well with a
thermal plasma emission model with a single absorption component. We
found no compelling need for an additional non-thermal high-energy
component. We discuss these results in the context of a currently
favored models for $\gamma$ Cas and its { \it analogs}.

\end{abstract}



\section{INTRODUCTION}
Classical Be stars are B stars whose spectra contain emission lines
from H I while also showing evidence for rapid rotation and
equatorially concentrated circumstellar disks [see
  \cite{Rivinius2013} for a recent review of Be stars]. $\gamma$ Cas,
also known as HD~5394, was the first known Be star \citep{secchi67}.
It is of spectral type B0.5 IVe \citep{toko97} and is at a distance of
188 pc \citep{ci98} with a mass that has been estimated to be in the
range 13-18 M$_\odot$ \citep{zfc05, harm00, gat00, toko97}.  In
addition to being the prototypical member of its subclass and one of
the brightest, it has been source of extensive study due to its
unexpected X-ray properties.

$\gamma$ Cas has previously been studied in the X-ray band and found
to have luminosities in the range $10^{32} - 10^{33}\ {\rm
  ergs\ s^{-1}}$ and $L_{x}/L_{bol}>10^{-6}$ \citep[e.g.][]{smith2012,
  smith04, owen99}.  This is significantly larger than the typical
$\sim 10^{30}\ {\rm ergs\ s^{-1}}$ X-ray luminosity for single O and B
stars \citep{Naze2009, Oskinova2005, hm01, pall81} yet smaller than
the X-ray luminosity of Be-neutron star binary systems which during
quiescence typically exceed $10^{34}\ {\rm ergs\ s^{-1}}$ and during
outburst can exceed $\sim10^{38}\ {\rm ergs\ s^{-1}}$
\citep[e.g.][]{negu98}. There are two classes of models for producing
the observed X-rays in $\gamma$ Cas. The currently favored one invokes
a magnetic star-circumstellar disk interactions \citep{rsh2002} while
others involve accretion onto a putative white dwarf binary
companion. There are now as many as 10, and possibly 12, similar stars
identified in the literature e.g. \citep{nebot2013,rauw2013,
  Motch2007, Lopes2006, Safi-Harb2007} sometimes called $\gamma$ Cas
analogs. Each are Be stars of similar spectral type (B0.5 - B1.5) with
similar X-ray luminosities.  Three are apparently in clusters and it
has been suggested that they may be blue stragglers which is another
possible clue as to their underlying nature \citep{marco2007}.

The soft X-ray spectrum (below $\sim$12 keV) of $\gamma$ Cas is
thermal in nature, being well approximated by optically thin thermal
plasma models \citep{smith2012, smith04,owen99,kmic98,src98} and it
apparently consists of multiple components \citep{smith12,
  Lopes_de_Oliveira2010}. Above 10 keV, it is roughly similar in terms
of luminosity and spectral energy distribution to hard-X-ray selected
cataclysmic variables (CVs) which are primarily intermediate polars
\citep{Landi2009, Revnivtsev2008}. The presence of a white dwarf
binary companion has often been suggested as the source of the X-ray
emission.  However, those objects typically exhibit significant flux
and spectral variability above $\sim$20 keV on measurable timescales,
e.g.  \cite{Scaringi2010, Landi2009}. Furthermore, there is no
compelling evidence in $\gamma$ Cas for the presence of the
periodicities that are predominant in CV power density spectra.

Recently, \cite{torrejon2012} have presented a comprehensive
broad-band X-ray spectral study of the $\gamma$-Cas analog HD 110432;
also see \cite{deOliveira07}. They found that multiple temperature
components, possibly including a very hot thermal or non-thermal
powerlaw, are needed to adequately fit their data. This result helped
motivate us to further explore the brighter and more extensively
studied $\gamma$ Cas and in particular to make use of the
unprecedented body of high-energy observational data now available
from {\it Suzaku} and {\it INTEGRAL}. In particular, the Cassiopeia
sky region received extensive \INTEGRAL\ exposure during the first 8
years of the mission driven largely by nuclear astrophysics studies of
the Cas-A remnant.

Previous temporal analysis of the X-ray emission from $\gamma$ Cas has
revealed the presence of two components \citep{src98}: 1) a ``basal"
component that varies on timescales of hours and contributes
$\sim$70\% of the X-ray flux, and 2) a ``shot" component characterized
by fluctuations on timescales of seconds to minutes.  \cite{henry2012}
recently established a revised and more accurate value for the optical
signal modulation that they attribute to the stellar rotation
period. Their new value is $1.215811\pm0.000030$ days, obtained using
a comprehensive photometric history spanning a 14-year baseline.  If
the high-energy emission is, for example, associated with putative
magnetic regions on the Be star modulation at this period could
occur. However, that would require a field topology that is stable on
much longer time scales than the period.  $\gamma$ Cas is also known
to have a binary companion in a 203.55-day nearly circular orbit($e <
0.03$, e.g. \citep{Nemravova2012}, which is considered unlikely to be
associated with the X-ray emission \citep{smith2012}. We note that
Henry \& Smith (2012) can rule out photometric modulation at the 
binary period at the level of 5 mmag.

Other variability patterns are evident as well.  \cite{henry2012} have
identified cycles of amplitudes $\sim$0.02-0.03 mag with typical 2-3
month durations. These are in turn superimposed on the smaller
amplitude 1.21-day signal attributed to the stellar rotation
period. Color dependence was also evident as the V -band cycles to be
30\% - 40\% larger than the B-band cycles.  Corresponding variability
patterns in the high-energy emission could reveal deeper insight into
its origin.  Thus, searches for periodic and/or stochastic variations
in our data were made.

In this paper we consider the hard-X-ray properties of $\gamma$ Cas in
the context of this emerging multi-wavelength picture. Our goals are
to address the flux history and possible spectral evolution on a range
of timescales and interpret our results in the context of current
models for the high-energy emission. Of particular interest is the
presence, or not, of a non-thermal high-energy component.  In Section
2 we discuss our data collection and basic analysis.  In Section 3 we
present results of our imaging, temporal, and spectral analyses. We
consider the high-energy spectral energy distribution in greater
detail, combining the {\it INTEGRAL} data with X-ray data obtained with
{\it Suzaku}, and in particular offer arguments in support of our
assertion that in addition to the broad thermal X-ray continuum, a
high-energy component is not required to fit the data. In Section 4 we
discuss the possible implications of this spectral component along
with the apparent lack of high-energy variability and offer our
conclusions.

\section{OBSERVATIONS AND DATA ANALYSIS}

In this paper we present the results of the accumulated {\it INTEGRAL}
hard X-ray/$\gamma$-ray observations of $\gamma$ Cas that span about 8
years. In addition, we have obtained 52 ksec of {\it Suzaku}
observations with XIS and HXD instruments providing a high-quality
spectral determination over the nominal range of $0.1 - 30$~keV.

\subsection{INTEGRAL}

The {\it INTEGRAL} imager, IBIS, is a coded mask instrument which has
a wide field of view (FOV) of 29$^{\circ}$ $\times$ 29$^{\circ}$
(9$^{\circ}$ $\times$ 9$^{\circ}$ fully coded) with a point spread
function (PSF) of 12$'$ (FWHM) and is sensitive over the energy range
15 keV to 10 MeV.  For a complete description of the {\it INTEGRAL}
spacecraft and mission refer to \citet{wink03}.  There are two IBIS
detector layers: ISGRI, an upper CdTe layer with peak sensitivity
between 15 and 200 keV, and PICsIT, a bottom CsI layer, with a peak
sensitivity above 200 keV \citep{uber03}. In this paper we have used
only ISGRI data.

We have analyzed the public data from all spacecraft pointings, or
"Science Windows", herein ``SCWs'', obtained between December 2002 and
February 2011 for which $\gamma$ Cas (J2000 $\alpha$ = 14.177,
$\delta$ = 60.717) was within $10^{\circ}$ of the pointing direction.
The resulting dataset consisted of 2610 SCWs.  Data reduction was
performed using the standard OSA 9.0 analysis software package
available from the {\it INTEGRAL} Science Data Center.  Spectral and
timing analyses were performed using the XSPEC and XRONOS data
analysis packages, respectively.

\subsection{Suzaku}
The {\it Suzaku} satellite \citep{Mitsuda2007} observed $\gamma$ Cas
on 2011 July 13.  During our {\it Suzaku} observations (obsid:
406040010), two sets of instruments were used: the X-ray Imaging
Spectrometer \citep[XIS,][]{Koyama2007} on the focal plane of the
thin-foil X-Ray Telescope \citep[XRT,][]{Serlemitsos2007} and the Hard
X-ray Detector \citep[HXD,][]{Takahashi2007,Kokubun2007}.  These
instruments operate simultaneously on every observation: the net
exposures of the XIS and the HXD are 55.4~ksec and 52.3~ksec,
respectively.  A detailed description of the {\it Suzaku} spacecraft
and mission can be found in the {\it Suzaku} technical
description\footnote{http://heasarc.nasa.gov/docs/Suzaku/prop\_tools/Suzaku\_td/}.

The XIS comprises four X-ray CCD cameras, XIS0$-$3, three of which
(XIS0, 2 and 3) use front-illuminated (FI) CCD chips and one (XIS1)
does the back-illuminated (BI) chip.  The FI has good harder X-ray
sensitivity, covering $\sim$0.5$-$10~keV, while the BI has sensitivity
down to $\sim$0.3~keV. The XIS2 was fatally damaged on 2006 Nov 9 by a
micro-meteorite, so that during the observations, three instruments
(XIS0, 1, 3) were in operation. Each CCD was operated using the Spaced
Charge Injection (SCI) technique, which improves the spectral
resolution with a sacrifice of the effective imaging area.  The XRT
has a butterfly-shaped point spread function (PSF) with the half power
diameter (HPD) of $\sim$2$'$.
%
%
%
The HXD comprises two types of detectors, the PIN with the sensitivity
between 15$-$70~keV and the GSO between 40$-$600~keV.  Since $\gamma$
Cas is not bright enough between 40$-$600~keV for the GSO, we only
analyzed the PIN data.  The PIN detector has a collimator with a
34\ARCMIN$\times$34\ARCMIN\ \FOV, on the bottom of which are PIN Si
diodes installed.  There is no other known hard X-ray source in the
PIN \FOV\ nor is any seen in the {\it INTEGRAL} image that could
contaminate to the $\gamma$ Cas spectrum.

$\gamma$ Cas was put at the XIS on-axis (XIS nominal) position.  The XIS
was operated with the normal mode (no window option) because the count
rate of $\gamma$ Cas on each XIS, $\sim$6.6~\UNITCPS, was lower than
the pile-up threshold.  However, the XIS pileup estimator
\citep{Yamada2012} derived small pile-up of $\sim$3\% at the PSF core,
which flattens the XIS spectrum above $\sim$9~keV.  We therefore
excluded XIS spectra above 9~keV for these observations. In the data
analysis, we used the HEASoft version 6.15 and the CALDB version HXD
(20110913), XIS (20130916), XRT (20110630).

\section{RESULTS}

\subsection{INTEGRAL Imaging Analysis}

We used the OSA 9.0 software to produce mosaic images of the $\gamma$
Cas region in 3 broad energy bands: 20-40 keV, 40-60 keV, and 60-100
keV.  In Figure 1 we show the 20-40 keV and 40-60 keV significance
maps including all 2601 science windows.  They clearly indicate an
unambiguous detection of the source and its association with $\gamma$
Cas. The detection significance using the full dataset is 63.5$\sigma$
and 13.3$\sigma$ in these hard bands.  It is thus evident that $\gamma$ Cas
has a relatively steep spectrum in this energy range.

Mosaic maps from subsets of the total dataset were inspected to
determine a list of sources to include during the spectral and timing
analysis steps.  This was done to ensure that low duty cycle,
transient sources would be accounted for. The datasets consisted of all
SCWs from 100 orbit blocks of data. i.e. from orbits 1-100, 101-200,
etc.  We found 16  sources in the $\sim 20^0 X 20^o$ region around $\gamma$ Cas
that we detected at the 5$\sigma$ level or greater during at least
one subset. These sources, that are labeled in Fig. 1, were included
in our coded mask solution, leading to the background subtracted image
of Fig. 1, as well as in our and spectral extraction procedure. Fourteen
of these other sources are associated with previously known X-ray 
sources and the two new IGR sources have subsequently been associated with
a CV (IGR~J00234+6141) and a high-mass X-ray binary (IGR~J00370+6122)
\citep{bird2010}. 

\subsection{INTEGRAL Timing Analysis }

As noted earlier, $\gamma$ Cas has been shown to exhibit photometric
variations with a 203.55 day periodicity which has been speculatively
associated with the orbital motion of a $\sim$1 M$_{\odot}$ companion
in a near circular orbit \citep{harm00}.  In addition, a $\sim$1.2-day
modulation has been seen in the B- and V-band photometric monitoring
of \cite{henry2012}, also see \cite{smith06, harm00}, which is
believed to be due to the rotation of the Be star. Previous studies
have suggested the possible association of a $\sim$1-day modulation
with a white dwarf companion \citep{appa02}, although that was based
on the inaccurate period determination of \cite{src98} and the stellar
rotation interpretation is now favored.

The {\it INTEGRAL} data set is perfectly suited to produce a long-term
lightcurve in order to search for modulations associated with
these periodicities and other possible periodicities as well as for
stochastic variability. If, for example, the hard-X-ray emission is
associated with particle acceleration resulting from a magnetic
disk-star interaction scenario, it is not unreasonable to expect the
1.21-day period to emerge, or at least variations at comparable
timescales.

We constructed light curves based on single science windows flux
determinations in the 20-40~keV and 40-60~keV energy intervals. We
then binned the resulting series into sample sizes appropriate for a
general variability searches and for period searches, focusing on the
published X-ray and optical-UV period determinations.

The flux history of $\gamma$ Cas is shown in Figure 2 where we show
the 20-40 keV, 40-60 keV, and ratio light curves rebinned into 48 hour
time bins from the intrinsic $\sim$2000s single-science window time
samples (which defines our maximum time resolution) reflecting the
typical length of an {\it INTEGRAL} pointing. We also present the
hardness-intensity diagram derived from these measurements in the
right hand panel. From Figure 2 we can see by visual inspection that
there is no evidence for large flares but, since some time periods are
sparsely sampled, the possibility of low-amplitude flaring cannot be
ruled out.  We thus formally tested whether light curves, from this
time series as well as with several more coarsely binned ones, were
consistent with being constant using the chi-square statistic and
computing the mean and expected variance for the assumption of a
steady source.  We did this for different ranges of timescales by
binning the data into relatively short (0.5-day) and long (30-day)
time intervals.  The results in all cases were consistent within
statistics to the light curves being derived from a constant intensity
source.  For example, for 0.5-day bins using the higher
signal-to-noise 20-40~keV band, the observed variance was about one
third of the mean and equivalent to the expected variance within about
20\%. The corresponding null-hypothesis probability was $10^{-4}$.

We also performed periodicity searches and general tests for
variability on the {\it INTEGRAL} light curves. We computed the
power-density spectra (PDS) for a variety of cases, using the
20-40-keV light curve and the 20-60-keV light curves.  We also tried
this for the single SCW case, i.e. our maximum time resolutions and
for light curves binned at e.g. 8000-s, 1- and 2-day intervals. In
none of the resulting PDS did we find excess power at a statistical
significance of greater than 2$\sigma$ between the minimum resolvable
timescales and $\sim$100's of days. 

Additional variability searches were accomplished using a chi-square
test on epoch-folded light curves at and around known periods of the
system; specifically the 1.21-day and the 202-day periods associated
with Be-star rotation and companion-orbital frequencies. We
searched several hundred possible periods on intervals about those
values and performed a chi-square test for each case, again testing
the constant-source null hypothesis. The results were all
negative. For example, we used 20-40-keV band 10-phase light-curve
bins to search for 128 possible periods centered on 203.55 days. We
found the data to be consistent with having such a 203.55-day periodic
component to be only at the 41\% level of confidence. For a
1.215767-day period search the corresponding confidence level was only
9\%. No periodicities were found in the data above a level of 81\%.

\subsection{\SUZAKU\ Timing Analysis} 

As noted in Section 1, soft X-ray variability on multiple timescales
has been well documented. Here we examined our 56-ksec
\Suzaku\ integration to search for similar or other unanticipated
temporal behavior.

The XIS image shows only one very bright X-ray source, which is
$\gamma$~Cas.  In the XIS data analysis, we defined a source region
with a 3\ARCMIN\ radius circle centered at $\gamma$~Cas, which
includes $\sim$95\% of X-ray photons from the star.  We extracted the
background from an annulus region with a 5\ARCMIN\ inner radius and a
7.5\ARCMIN\ outer radius centered at $\gamma$~Cas.  The HXD/PIN data
include significant contamination from non X-ray background (NXB) and
cosmic ray background (CXB).  The NXB can be estimated at 1.3\%
uncertainty (1$\sigma$) using the tuned background model
(JX-ISAS-SUZAKU-MEMO-2007-09\footnote{ftp://legacy.gsfc.nasa.gov/Suzaku/data/background/pinnxb\_ver2.0\_tuned/}).
The CXB is estimated from the typical CXB emission \citep{Boldt1987},
which may fluctuate by $\lesssim$30\% from region to region
\citep{Miyaji1998}.

We produced light curves of the XIS FI sensors (XIS0 and XIS3) between
4$-$9~keV and the HXD/PIN light curves between 15$-$40~keV.  Each XIS
sensor collected $\sim$0.8~\UNITCPS\, while the HXD/PIN collected
0.16~\UNITCPS.  The combined XIS light curve showed strong spiky
variation by $\approx$50\% on timescales of $\sim$1~ksec
(Figure~\ref{fig:SuzakuLC}), which is similar to what was observed
with \XMM\ \citep{Lopes_de_Oliveira2010}.  We also depict the
hardness-intensity space behavior based on the same data in the right
hand panel of that figure. The flux went down gradually during the
observation, reached the lowest flux at $\sim$10$^{5}$~sec after the
observation start, then recovered to the original flux level. We note
that given the relatively short duration of the observation, compared
to the \INTEGRAL\ data set we could not address the presence or
absence of the reported optical-UV periodicities in this band. We
found that the variation fraction in the \SUZAKU\ data on 5 ksec
timescale is less than a factor of 2.

Since the background uncertainty is more significant in the HXD data,
the HXD/PIN light curve generated with the HEASoft tool {\tt
  hxdpinxblc} only includes bins with fractional exposures more than
99\%.  Therefore, some bins in the XIS light curve are not present in
the HXD/PIN light curve.  The HXD/PIN count rate was nearly constant
for the first $\sim$7$\times$10$^{4}$~sec with a small dip at
5$\times$10$^{4}$~sec. This was followed by a steep decline by a
factor of 3 between 7$\times$10$^{4}$~sec and $\sim$10$^{5}$~sec, and
then recovery to the original level by the end of the observation.
The variation was very similar to that in the XIS 4$-$9~keV band.  The
hardness ratio (HR) defined by the ratio of the HXD/PIN 15$-$40~keV count
rate over the XIS 4$-$9~keV count rate (bottom panel of
Figure~\ref{fig:SuzakuLC}) does not show any remarkable variation.
The HR curve can be accepted by a fit with a constant model at above
90\% confidence (reduced $\chi^{2}$ = 1.16, d.o.f. = 83).

\subsection{\INTEGRAL\ Spectral Analysis}

As a first step to explore the \INTEGRAL\ data set we modeled this
spectrum independently of the \SUZAKU\ data.

We extracted the IBIS/ISGRI hard X-ray time-integrated spectrum for
the entire 8-year span using the \INTEGRAL\ OSA tools. Previous
analyses of broadband X-ray spectra of $\gamma$ Cas have generally
found that the spectra were thermal in nature, typically modeled by
optically-thin, thermal plasmas, \citep{smith2012, owen99,
  src98, kmic98, hora94, mura86} but in some cases a power-law (or
cut-off power-law) model plus a Gaussian line also provided an
adequate fit \citep{hart06, src98, kmic98,pisw93}. We also note the
recent analysis of the $\gamma$ Cas analog HD 110432 by
\cite{torrejon2012}, where the possible presence of a hard,
non-thermal spectral component is identified.

A primary objective here is to characterize the shape of {\it
  INTEGRAL} spectrum and given the apparently flat hard X-ray light
curves we focused on the time-integrated dataset.  This will offer the
best opportunity for the foreseeable future to explore possible
deviations from the broad-thermal spectral forms currently
established. Towards this end, we first fit the $20-100$~keV continuum
using a cutoff powerlaw form. Integrated over the 20-100-keV band our
best fit model corresponds to a flux of 4.9$\times 10^{-11}$
ergs/cm$^2$/sec. In order to best test the hypothesis that an
additional hard component, perhaps non-thermal in nature, is 
superposed on the broad thermal emission we first fitted models to the
\INTEGRAL\ data alone.  We also tried thermal Comptonization spectral
forms. In all of these scenarios we found evidence of positive
residuals above $\sim 40$~keV. We thus concluded that the addition of
a hard powerlaw component, presumably representing a non-thermal
contribution to spectrum, led to a nominal improvement in the quality
of the fit. However, an F-Test applied to the two results
indicated a probability of only 0.42 that the cutoff powerlaw / thermal
Comptonization model fits were improved through the inclusion of the
hard tail. Furthermore, that result was strongly dependent on the highest
energy channels where the data becomes signal-to-noise limited, and
the powerlaw component contributes only about 5\% of the flux.  In any
case our objective was to assess how this scenario would hold up to
the further scrutiny of fitting the combined broad band data sets, for
which a global continuum model can be more meaningfully constrained.

\subsection{\SUZAKU\ Spectral Analysis}

$\gamma$ Cas has been observed by a number of current major X-ray
observatories covering the 0.2-10\,keV domain including both Chandra
and XMM-Newton.  Its X-ray spectrum consists of a strong thermal
continuum with $kT\approx 12-15$\ keV, plus some broad ($\approx
600$\,km\,$^{-1}$) and symmetric X-ray lines which have been modelled
by optically thin plasmas {\bf (e.g. \cite{smith2012} }. Models have
included as many as 4 components, with temperatures ranging from 0.15
to 12 keV, and fluorescence K features from iron and silicon. It
should be noted though that the hottest of these putative gas phases
is typically the predominant component of the emission contributing
more than the combined cooler phases.

In the context of the photoionized plasma models, different iron
abundances were inferred between the measurements with the K-shell
line that originates from $\sim$140 MK plasma and Fe L-shell line
complex from the 1-6 MK plasma.  Smith et.~al.~(2004) suggest that
these harder X-rays are seen through the dense regions of the
circumstellar disk, while most of the X-ray emission is absorbed only
by the stellar wind or the outer, less dense regions of the disk. The
presence of fluorescence features indeed suggests the presence of gas
close to the X-ray emission regions likely associated with the
circumstellar medium.  These findings are confirmed by analysis of RGS
XMM-Newton spectra by Lopes de Oliveira (2010). Those authors report
FeXXV and Fe XXVI Ly K line strengths consistent with their inferred
12-14-keV plasma temperature, but with a subsolar Fe abundance of
about 0.2 to 0.4.  This is puzzling because the chemical abundances of
typical B stars within 1 kpc of the Sun are consistent with solar
values \citep{Przybilla2008}.

Figure~\ref{fig:SuzakuSpec} shows the XIS0+3 and HXD/PIN spectra of
$\gamma$~Cas above 0.6 keV.  The spectrum below 10~keV is similar to
those in the earlier \CHANDRA\ and \XMM\ observations: hard spectra
with three weak emission lines from Hydrogen-like and Helium-like iron
and iron fluorescence \citep{smith04,Lopes_de_Oliveira2010,smith12}.
The HXD/PIN spectrum clearly extends up to $\sim$40~keV.

\citet{smith12} discussed the presence of a
deeply embedded (\NH~$\sim$7$\times$10$^{23}$~\UNITNH) hard X-ray
component, whose continuum emission is $\sim$1/10 of the observed
spectrum in the hard energy band.  They discussed that an optically
thick ejecta from $\gamma$ Cas may block a part of the hard X-ray
emission, but the Compton reflection of the incident X-ray emission at
a surface, which should be accompanied by the iron fluorescence line
at 6.4 keV, can produce such a deeply embedded hard X-ray spectrum.
We briefly estimated this contribution using the {\tt pexmon} model in
{\tt xspec}, then the reflection continuum flux can be as high as
$\sim$1/10 of the direct emission from $\gamma$~Cas around $\sim$6~keV
when the elemental abundance of the reflecting material at
$\sim$0.32~solar, viewing angle of $\sim$30\DEGREE\ and the reflection
scaling factor of $\sim$0.3, for example.

Because this reflection component contributes up to a few tens of
percent at the energies of interest, we fit the spectrum by a
bremsstrahlung model for the continuum and 5 Gaussians for the iron,
sulfur and neon emission lines.  In this model fit, the model
normalization for the HXD/PIN spectra were multiplied by 1.15,
according to the Suzaku Data Reduction
Guide\footnote{http://heasarc.gsfc.nasa.gov/docs/Suzaku/analysis/abc/}.
The best-fit model (Table~\ref{tbl:specfit_suzaku} and
Figure~\ref{fig:SuzakuSpec}) reproduced the spectra well (reduced
chi-square of $\chi^2$ =1.04, d.o.f =890).  The derived plasma
temperature, \KT~$\sim$14.4$\pm$0.4~keV, is similar to those in
earlier observations \citep{smith04,smith12}.  The EWs of the iron
emission lines are also similar to those measured with
\CHANDRA\ \citep{smith04}, but a factor of two higher than those in
the {\XMM} observation in 2004 \citep{Lopes_de_Oliveira2010}.

\subsection{Combined Spectral Analysis}

The hardness ratios of both the \SUZAKU\ 5$-$9~keV/15$-$40~keV bands
and \INTEGRAL\ 20$-$40~keV/40$-$60~keV bands did not show any
significant temporal variations.  This suggests that the \SUZAKU\ and
\INTEGRAL\ time-integrated spectra can be jointly fitted by
normalizing them at the overlapping energy band between 20$-$40~keV.

For the combined data set we found that the collisionally ionized
plasma emission model, {\it apec} in the XSPEC package, along with a
single absorption term and a Gaussian line component at $6.4$ keV
provided an excellent fit to the full data set as was the case with
the \Suzaku\ data only.  The inferred plasma temperature was about
$15$ keV, comparable to what others have found for $\gamma$ Cas
(e.g. \cite{smith04}) and less than that the hot component of
HD~110432 of \cite{torrejon2012}. Significantly, the combined fit
provides strong evidence that the thermal emission extends up to $\sim
100$keV. The inferred luminosity over the $0.1-100 $keV band is
$log(L_x) = 32.4$. Surprisingly, the continuum extended smoothly into
the higher energy \INTEGRAL\ band without requiring a significant
instrumental cross-calibration correction. The reduced chi-square
statistic for our best overall fit to that model was 1.28 for 1539
degrees of freedom.

We tried adding a powerlaw component to the model, and found that a
similar quality of fit could be obtained. However, that was the case
only when the hard powerlaw, index $\Gamma = 1.67$, made negligible
contribution to the overall flux ($< 1\%$). Formally, and more
dramatically than the \INTEGRAL\ only case, an F-test indicated only a
probability of $<1\%$ that the additional component led to a more
accurate representation of the data. Thus, we suggest that a purely
thermal plasma emission model is the most realistic representation of
the data and the addition of a separate hard component is not
justified by our observations. 

The best-fit results for our combined model are presented in 
(Table~\ref{tbl:specfit_INTEGRAL+Suzaku}).

\section{Discussion and Conclusions}

The ultimate question of how one resolves the long standing $\gamma$
Cas enigma remains but some new clues about this system are slowly
emerging. The broadband X-ray emission from the Be stars $\gamma$ Cas
can contribute to this endeavor.  There is, at this point in time, a
fairly broad consensus that the standard Be star X-ray binary model
does not provide a plausible description of $\gamma$ Cas. This is in
part the case given the lack of orbital modulation and episodic
variability as well as the discrepant X-ray luminosity compared to
quiescent Be X-ray binaries. Most notable from the perspective of this
paper is the nature of the high-energy continuum which in Be X-ray
binaries generally displays non-thermal emission components. In
transients this can be the case over a range of luminosities; for a
recent example see \cite{Kuhnel2013}.

Scenarios involving a white dwarf companion remain possible but have
long been considered problematic and no clear X-ray signatures of
binarity have been revealed. One difficulty however with the Be-white
dwarf binary scenario for these stars is the problem of forming such a
system given the estimated mass ranges for $\gamma$ Cas of 13-18
M$_\odot$ \citep{harm00, zfc05, gat00, toko97}. In order for a white
dwarf to form in a Be system, it had been long believed white dwarf
progenitor must initially be more massive than the Be star
progenitor. Evolutionary models by \citet{ragu01} suggest that the
maximum initial mass of a white dwarf progenitor in a Be binary system
is 10-11 M$_\odot$ although variations of the traditional shared
envelope binary evolution scenarios, involving for example
non-conservative mass transfer, could plausibly circumvent this. In
the context of this constraint $\gamma$ Cas would then have needed to
be initially $\leq$10 M$_\odot$ for a white dwarf companion to have
formed, but the accretion of 3 - 8 M$_\odot$ or more from the
companion to reach its current mass seems unlikely. A putative third
body could also alter this picture, but that scenario is highly
speculative and improbable.

From purely an observational perspective, while the
\INTEGRAL\ CV sample, e.g.  \citep{Landi2009, Revnivtsev2008}, is
characterized by a mean (log) luminosity of $L_{20-100keV} \simeq
32.5$ which is similar to $\gamma$ Cas, the short-timescale
variability characteristics are distinct. The INTEGRAL (and Swift)
CV sample is dominated by intermediate polars (IPs) characterized by
spin and orbital pulsations and aperiodic fluctuations, none of which
is evident here. The CVs typically exhibit softer spectra over that
band than what we find for $\gamma$ Cas, on average $\Gamma$ $\simeq
2.9$, however CV spectra vary significantly in both a phase dependent and
an episodic manner.

While an accreting neutron star scenario can be largely ruled out, we
point out a curious similarity between of $\gamma$ Cas and the
$\gamma$-ray binary LS~I+61~303, e.g. \citep{chernyakova2006}.  That
system, which consists of a similar B0Ve primary, but has a similar
mean 20-60~keV X-ray luminosity (although its 2-10~keV luminosity is
nearly an order of magnitude larger). It is believed, in one class of
models at least, to consist of a non-accreting companion, a
fast-spinning highly magnetized neutron star and the Be primary, thus
in essence a pulsar-wind nebula embedded in the Be star wind. This
interpretation requires that pulsations be cloaked by the massive star
wind.  Could the high-energy emission from $\gamma$ Cas also be
powered by a non-accreting companion? There are however clear
observational differences between these systems as well. LSI~+61 303
exhibits orbital modulation in X-rays, and is a radio source which is
also periodically modulated. Additionally, it is a $>$ 100-MeV
gamma-ray source \citep{nolan2012} as well as a TeV gamma-ray
source. $\gamma$ Cas on the other hand is apparently radio quiet, it
is not seen down to the 2.5 mJy flux limit of the NVSS for example
rendering this idea less tenable (although see \cite{drake1990} who
found a sub-5$\sigma$ (25$\pm$0.06~mJy) source coincident with
$\gamma$ Cas, so variability may be an issue. In any case its putative
mean radio luminosity would be $\sim10^3$ times lower than that of
LSI~+61~303 which is about 10 times more distant and 10 times
brighter).  To further explore this idea though we searched for
$>$100-MeV gamma-ray emission from $\gamma$ Cas, which is not listed as a
detection in the Fermi 2FGL source catalog \citep{nolan2012}, using
the archives of the Fermi Gamma-Ray Space Telescope. Our searches,
both deep, long baseline integrations and epoch-folding analyses on
the known periods revealed non detections. Thus, unless a scheme for
hiding radio and gamma-ray emission can be concocted, this idea
seems untenable.

A very different concept to explain the X-ray emission is the so
called ``magnetic star-disk interaction" model which was proposed by
e.g. \cite{smith04,rsh2002,rs00,sr99,src98}; also see
\cite{smith2012}.  In this model it is suggested that many
characteristics of the X-ray emission of $\gamma$~Cas can be explained
by the dynamical interactions between putative magnetic fields on the
star and its circumstellar disk.  This seems to be the favored
modeling scenario in the current literature, however there are still
some puzzling issues; not the least of which is the origin of the
magnetic field.

In this scenario short-timescale X-ray variations, as well as
correlated UV and optical activity is associated with matter in the
proximity of the star being acted upon by magnetic lines of force.
This mechanism involves a cyclical magnetic dynamo driven by a
Balbus$-$Hawley instability located within the circumstellar disk
\citep{rsh2002}. According to that model, variations in the magnetic
field strength emanating from the star lead to Keplerian shear and
turbulent motions within the disk through the magnetorotational
instability of Balbus \& Hawley (1991). That mechanism is expected to
operate to some extent whenever a magnetic field is embedded in a
Keplerian disk.  This could result in a positive feedback in which
turbulence amplifies the magnetic field, which, in turn, increases the
level of turbulence.

The more slowly rotating circumstellar disk drags the field
lines causing them on occasion to get entangled. The resulting
stretching and reconnection of the field lines leads to particle
acceleration, some of which may be directed towards the Be star
surface. These streams of plasma then locally heat the photosphere
leading to flaring. The heated matter can then expand to lower density
regions and continues to glow as the so called "basal" emission
component, e.g. \citep{src98}.

The model has been applied to the flux variations that were found on
several time scales utilizing optical, UV, and X-ray monitoring of
$\gamma$ Cas as detailed in \cite{smith04}.  It was discovered that
there was a cyclical variation in both the UV and X-ray data with a
1.21 day period but that the light curves in the two bands were
$180^{\circ}$ out of phase \citep{rs00,src98}. Rapid X-ray
fluctuations on timescales of seconds to minutes, termed ``shots'' or
``bursts'', have also been noted, e.g.  \cite{src98, rsh2002}.  The
number and amplitude of these shots were found to increase during the
UV minimum.  More recently a longer timescale was
identified. \citet{rsh2002} found that the optical and X-ray emission
from $\gamma$-Cas exhibited cyclical variations superimposed on the
1.21-day stellar rotation period with periods that changed from 55 to
93 days. This has subsequently been revised with improved measurement
to 50-91 days by \cite{henry2012}.  In this case the optical and X-ray
fluxes showed a strong positive correlation.

It was proposed that the shots were generated in magnetic structures
near the stellar surface and that the 1.21 day period is due to
rotation of the star \citep{rs00,src98}. A possible signature of such
models may be magnetically induced flaring, which would likely be
pronounced in the hard X-ray band (i.e. above 10~keV). It would seem
highly likely that the high-energy emission would be variable over
observable timescales in such scenarios. In the sun for example, where
we know for certain that magnetic reconnection events occur they are
known to cause particle acceleration leading to rapid and energetic
flaring. The disruption and reconfiguration of magnetic field
structures associated with the model could also provide a natural
explanation for a non-thermal high-energy emission component (although
it could also underlie the emergent thermal emission through radiative
transfer processes), but we fail to convincingly identify such a
component.

If magnetic fields emanating from the star, as opposed to from a
dynamo within the disk, are involved with the X-ray production the
question of how fields are generated in early-type stars is another
issue one must ultimately resolve. It has been suggested by some that
Be stars could be the end product of a merger event.  For example,
\cite{demink2013} suggest that these putative magnetic fields may play
a role in the binary evolution or merger events leading to the high
rotation rates seen in Be stars. If $\gamma$-Cas had gone through a
previous phase involving a relatively recent merger a remnant magnetic
field could exist at present. However, the presence of a companion
star, orbiting at the aforementioned 203-day period, would then
require the merger scenario to have occurred in a triple star system.

There is some observational evidence for large-scale magnetic fields
in at least a small fraction of O and B stars. Based on a survey of
550 Galactic O and B stars designed to search for stellar magnetic
fields \cite{Wade2013} find that about $7\pm1\%$ of their sample
objects are detected. However, their survey included 98 classical Be
stars, among which they failed to obtain any detections despite a
magnetic sensitivity similar to that of their larger sample.  Based on
the detection rate they measured for the non-Be B-type stars, those
authors expected to detect $10\pm2$ magnetic stars amongst the Be
stars. Thus, this seems to be a statistically significant result and
it argues against the notion that Be star circumstellar disks are
unlikely to be of magnetic origin.

There are some B-stars with spectral types similar to $\gamma$\ Cas
known to possess substantial magnetic fields. The X-ray properties of
these were studied in detail by \cite{Oskinova2011} . It was shown
that the X-ray properties of these massive B-type magnetic stars are
diverse.

An interesting case is an B0.5IV star $\xi^1$\,CMa whose spectral type
is the same as $\gamma$\,Cas. $\xi^1$\,CMa rotates somewhat slower
($P_{\rm rot}$=2\,d,) \cite{Hubrig2011} than $\gamma$\,Cas. It does not
exhibit characteristics of a disk and has an exceptionally strong, 5\,kG,
magnetic field. The detailed study of X-ray emission from $\xi^1$\,CMa
revealed X-ray pulsations coherent with optical pulsations but with a
larger amplitude \citep{Oskinova2014}.  However, despite the confirmed
presence of strong magnetic field, the X-ray spectrum of $\xi^1$\,CMa
is much softer than in $\gamma$\,Cas. Comparison between those two
objects shows that the very special X-ray properties of $\gamma$\,Cas
cannot be explained by only its putative magnetic field. The presence
of a disk or a hypothetical close companion is apparently required to
explain $\gamma$\,Cas and its analogs.

The lack of an obvious hard, non-thermal component in our analysis of
$\gamma Cas$, which might imply that particle acceleration is
occurring within the star-disk system, is consistent with the lack of
radio emission. However the star-disk interaction model details remain
to be worked out, although for now it seems to be the favored
explanation for the long-standing $\gamma$~Cas enigma.

To summarize, we have examined more than 8 years of non-uniformly but
reasonably well-sampled \INTEGRAL\ observations with fields of view
containing $\gamma$ Cas. We do not find significant flux variations in
\INTEGRAL\ band, either stochastic or periodic. Our frequency searches
included some known periods of the system. We also present our
analysis of a 52~ksec \SUZAKU\ observation. Temporal analysis of those
data considered with the \INTEGRAL\ analysis suggest a decreasing
variability with energy. Our multi-instrument spectral analysis
suggests a spectral energy distribution which is purely thermal in
nature over the broad $\sim 1-100$~keV band. While this result is
generally consistent with the favored magnetic field - circumstellar
disk interaction model it potentially imposes new constraints as the
magnetic reconnection mechanism that model invokes could quite
possibly entail both high-energy flaring and particle acceleration
leading to non-thermal emission.
 

\begin{acknowledgements}
{\bf ACKNOWLEDGEMENTS}

This project made use of observational data awarded to one of us (KH)
through the \Suzaku\ Guest Investigator program and \INTEGRAL\ archival
data obtained from the HEASARC at the NASA Goddard Space Flight
Center. CRS wishes to thank Myron Smith for many useful discussions
and for introducing him to this topic many years ago. LMO acknowledges
support from DLR grant 50 OR 1302.

\end{acknowledgements}

\bibliographystyle{apj}
\bibliography{inst,sci_AI,sci_JZ,scibook,bibfile.bib}


\begin{deluxetable}{llll}
\rotate
\tabletypesize{\scriptsize}
\tablecolumns{4}
\tablewidth{0pc}
\tablecaption{Suzaku Spectral Model \label{tbl:specfit_suzaku}}
\tablehead{
\colhead{Parameters}&
\colhead{unit}&
\colhead{Value}\\
}
\startdata
\multicolumn{2}{l}{Bremsstrahlung}\\
~~\KT\			&(keV)	&14.06~(13.81,14.30)\\
~~\EM\			&(10$^{55}$~cm$^{-3}$)&3.856~(3.841,3.869)\\
\multicolumn{2}{l}{Fluorescence iron line}\\
~~Line Center	&(keV)&6.405~(6.390,6.423)\\
~~$\sigma$		&(eV)&33~($<$74)\\
~~Flux			&(10$^{-5}$~ph~cm$^{-2}$~s$^{-1}$)&5.8~(4.8,6.9)\\
~~EW			&(eV)&52\\
\multicolumn{2}{l}{Helium-like iron line}\\
~~Line Center	&(keV)&6.680~(6.669,6.692)\\
~~$\sigma$		&(eV)&0~($<$40)\\
~~Flux			&(10$^{-5}$~ph~cm$^{-2}$~s$^{-1}$)&8.7~(7.8,9.4)\\
~~EW			&(eV)&84\\
\multicolumn{2}{l}{Hydrogen-like iron line}\\
~~Line Center	&(keV)&6.944~(6.930,6.959)\\
~~$\sigma$		&(eV)&0~($<$38)\\
~~Flux			&(10$^{-5}$~ph~cm$^{-2}$~s$^{-1}$)&5.9~(5.2,6.7)\\
~~EW			&(eV)&62\\
\multicolumn{2}{l}{Helium-like iron line}\\
~~Line Center	&(keV)&2.564~(2.542,2.586)\\
~~$\sigma$		&(eV)&100~(78,122)\\
~~Flux			&(10$^{-4}$~ph~cm$^{-2}$~s$^{-1}$)&1.4~(1.1,1.7)\\
~~EW			&(eV)&28.7\\
\multicolumn{2}{l}{Hydrogen-like iron line}\\
~~Line Center	&(keV)&1.013~(0.998,1.025)\\
~~$\sigma$		&(eV)&25~($<$48)\\
~~Flux			&(10$^{-4}$~ph~cm$^{-2}$~s$^{-1}$)&1.5~(1.0,2.1)\\
~~EW			&(eV)&8.6\\
\multicolumn{2}{l}{\NH}\\
~~\NH	&(10$^{20}$~cm$^{-2}$)&5.1~(4.7,5.4)\\
\enddata
\tablecomments{
}
\end{deluxetable}

\begin{deluxetable}{llll}
\rotate
\tabletypesize{\scriptsize}
\tablecolumns{4}
\tablewidth{0pc}
\tablecaption{Combined Spectral Model \label{tbl:specfit_INTEGRAL+Suzaku}}
\tablehead{
\colhead{Parameters}&
\colhead{unit}&
\colhead{Value}\\
}
\startdata
\multicolumn{2}{l}{Collisionally ionized thermal plasma (apec)}\\
~~\KT\			&(keV)	&15.37$\pm$0.18\\
~~Abundance		&solar  &0.26$\pm$0.18\\
~~Normalization	&       &7.76e-2$\pm$3.11e-4\\
\multicolumn{2}{l}{Gaussian}\\
~~Line Center	        &(keV)  &6.40 \\
~~$\sigma$		&(keV)  &7.49e-2$\pm$2.72e-2\\
~~Normalization	        &       &5.16e-5$\pm$7.99e-6\\
\multicolumn{2}{l}{Absorption (TBabs)}\\
~~NH     		&10$^{22}$cm$-2$  &2.56e-2$\pm$3.82e-3\\
\multicolumn{2}{l}{ChiSquare per DoF}\\
~~1709.14 / 1427	        &       &  \\
\enddata
\tablecomments{Example of a thermal plasma emission model fitted to
  the broad-band multi-instrument data set. The continuum appears to
  be well represented by a purely thermal from below a keV up to
  nearly 100 keV. An empirically determined instrumental cross
  calibration correction term was included in the model fit.
}
\end{deluxetable}

\begin{figure}[h]
\epsscale{1.0}
 \plotone{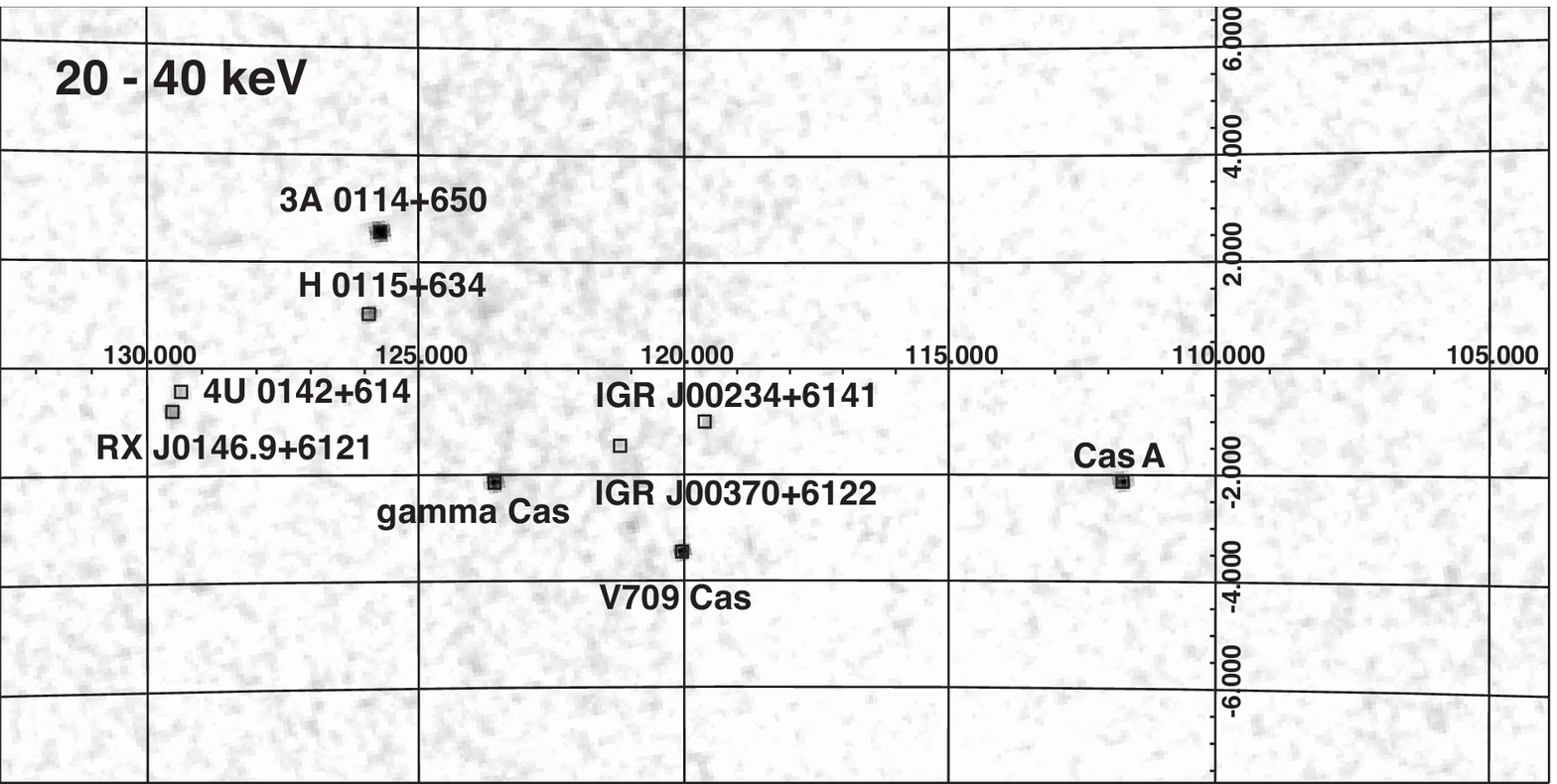}
 \plotone{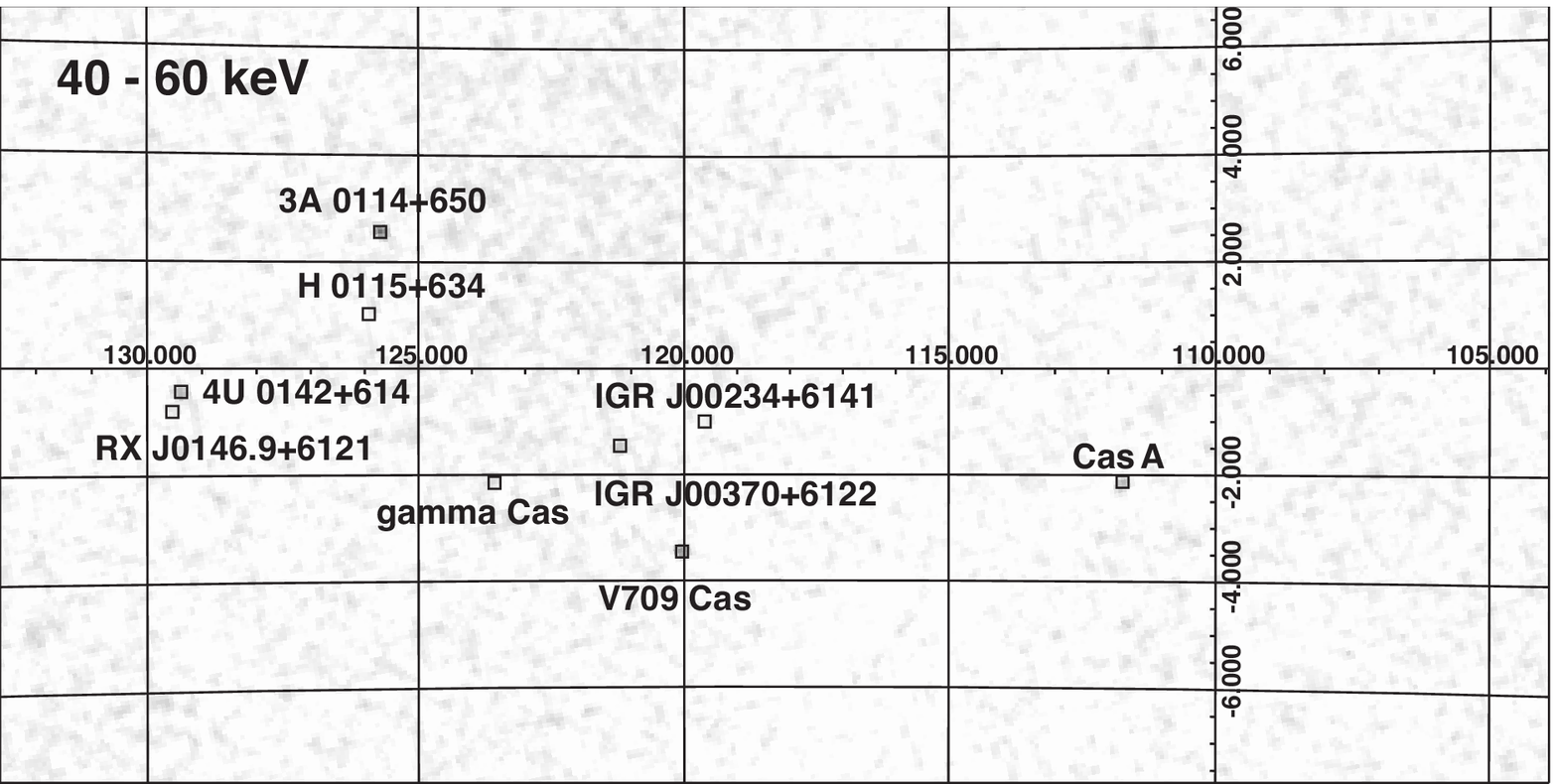}
\caption{The IBIS/ISGRI mosaic significance maps for
$\gamma$ Cas in the 20 - 40 keV and 40 - 60 keV energy bands.  Note
how $\gamma$
 Cas dims significantly between the two energy bands.  The significance
 drops from 63.5$\sigma$ in the 20 - 40 keV band
 to 13.3$\sigma$ in the 40 - 60 keV band.  This is also reflected in the
 spectra shown in Figures 3 \& 4.}
\end{figure}

\begin{figure}[h]
\epsscale{1.0}
\plotone{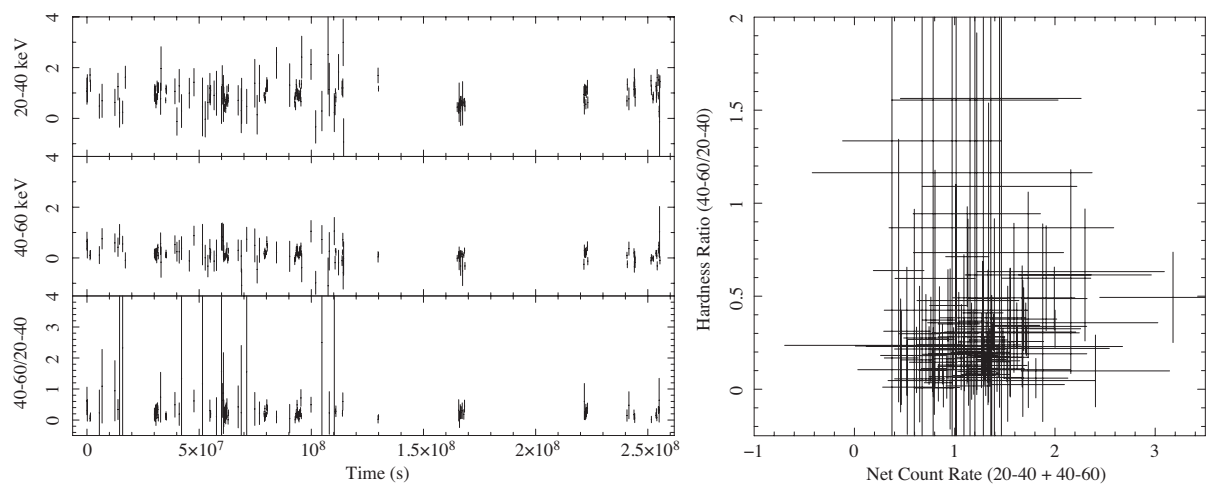}
 \caption{{\it Top and Middle left}IBIS/ISGRI
light curves for $\gamma$ Cas in the 20 - 40 and 40 - 60 keV energy
bands with 1$\sigma$ uncertainties.  The light curve consists of 2-day
time bins and covers the time period December 2002 to February
2011. The average count rate for this time period is 0.763 cts/s
$\sim$6.0 mCrab.  For reference, t=0 is MJD = 55755.00035. The
right hand panel depicts the hardness intensity space behavior based
on these light curve measuremnts.}
\end{figure}

\begin{figure}\epsscale{1.0} 
\epsscale{1.0}
\plotone{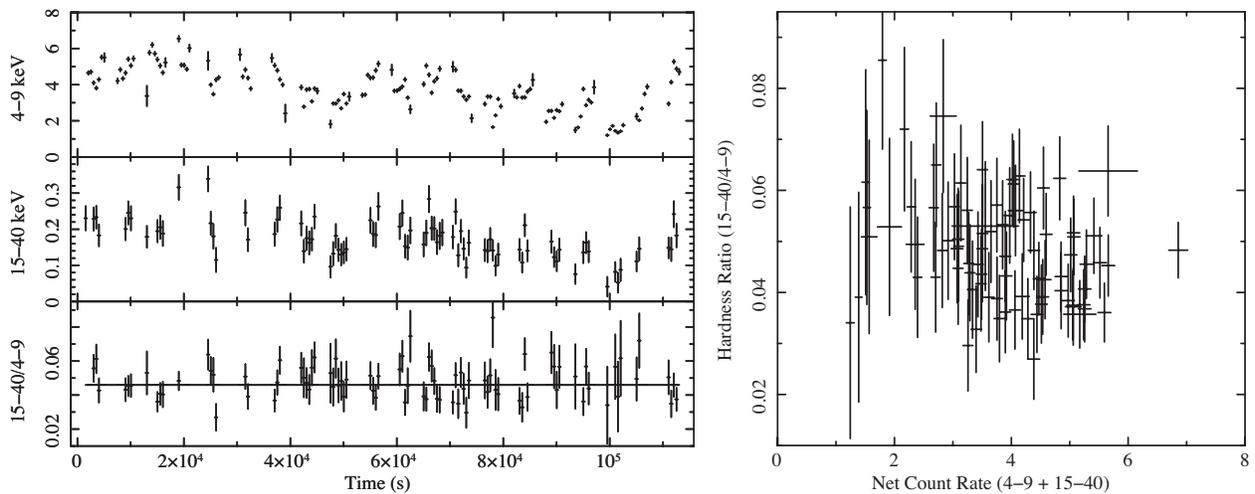}
 \caption{{\it Top left}: the background subtracted XIS (0+1+3) light curve
   of $\gamma$ Cas between 4$-$9~keV.  {\it Middle}: the background
   subtracted HXD/PIN light curve between 15$-$40~keV.  {\it bottom}:
   the hardness ratio of these two bands defined by the ratio of the
   HXD/PIN count rate over the XIS 4$-$9 keV count rate. The
   right hand panel is a hardness intensity diagram based on the same
   measurements. To within statistics, the hardness radio is constant
   over the factor of $\sim$3 intensity amplitude variation.
 \label{fig:SuzakuLC}}
\end{figure}

\begin{figure}\epsscale{1.0} 
\epsscale{1.0}
\plotone{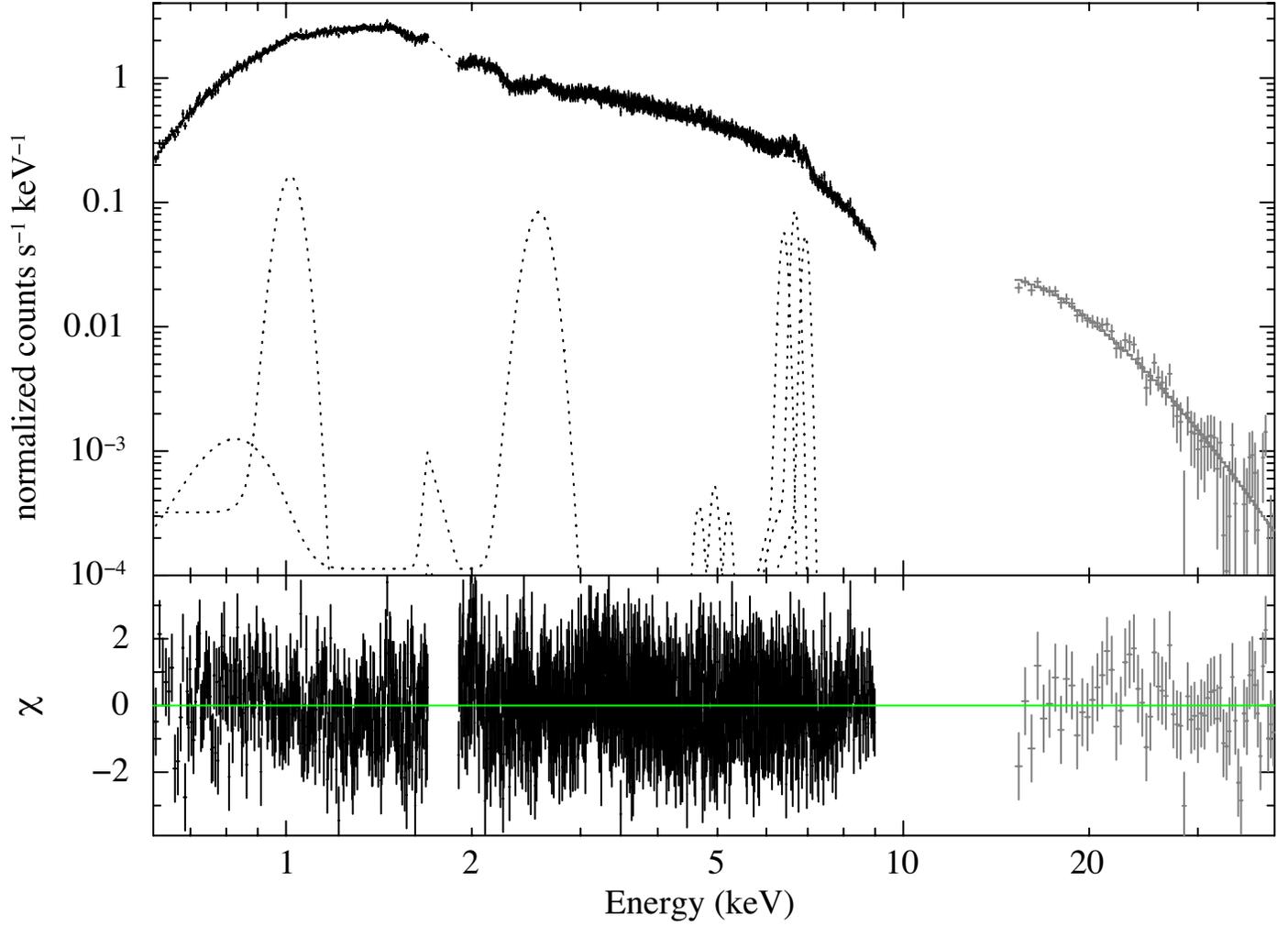}
\caption{\SUZAKU\ spectra of $\gamma$~Cas with the XIS0+3 ({\it black}) 
and the HXD/PIN ({\it grey}).
The solid line shows the bremsstrahlung model with 5 Gaussian lines,
and the dotted lines show individual components.
\label{fig:SuzakuSpec}}
\end{figure}

\clearpage 
\begin{figure} 
\epsscale{1.0}
\centering
\includegraphics[totalheight=0.045\textheight,angle=90]{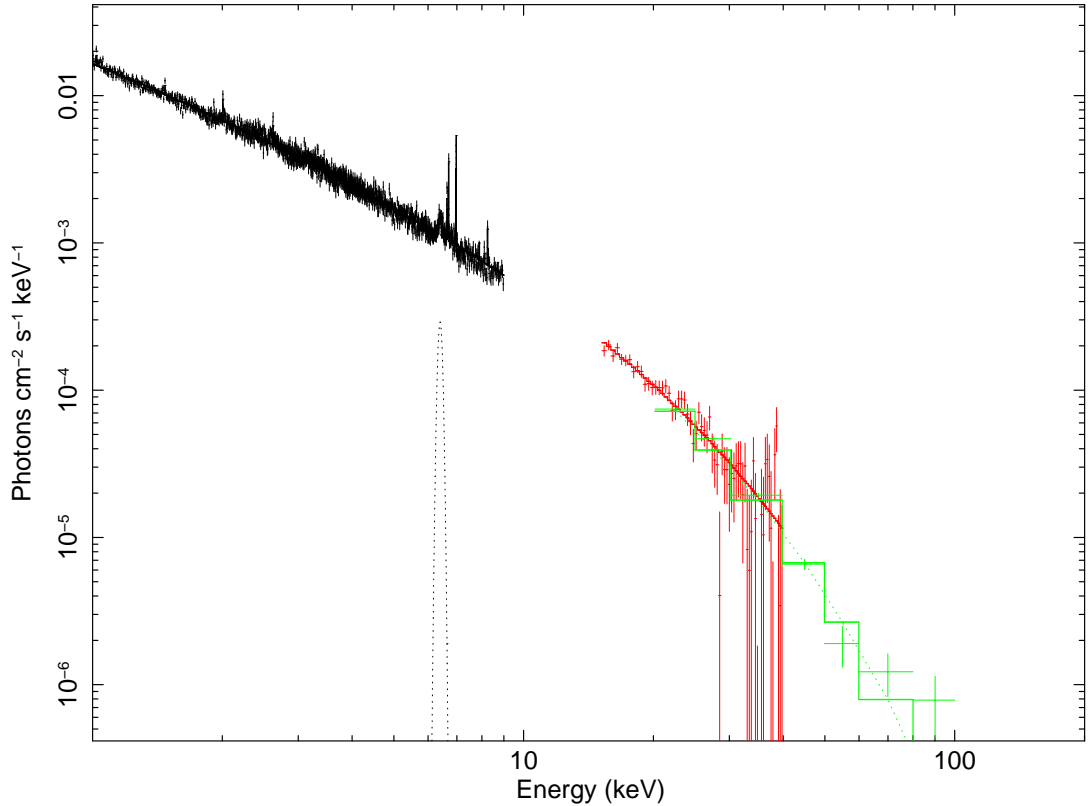}
\vspace{2.5 in}
\caption{
Combined broad-band time, integrated high-energy spectrum of
$\gamma$~Cas derived from our analysis of the \Suzaku\ and
\INTEGRAL\ datasets described in the text. The model consists of
emission from collisionally ionized plasma with temperature of 15.3
keV (based on the ``apec'' model of the XSPEC software), a Gaussion
line feature at 6.4 keV and a single absorption component. Negligible
correction for instrumental cross calibration was
required. Here we fit the complete data set
from $\sim$0.6-100-keV. Clearly, the model represents the data
extremely well. }

\end{figure}

\end{document}